\documentclass[usenatbib]{mn2e}
\usepackage{epsfig}
\usepackage{subfigure}
\usepackage{amssymb}
\usepackage{color}
\newcommand{\bn}{\begin{enumerate}}
\newcommand{\en}{\end{enumerate}}
\newcommand{\bi}{\begin{itemize}}
\newcommand{\ei}{\end{itemize}}

\def\gtorder{\mathrel{\raise.3ex\hbox{$>$}\mkern-14mu
    \lower0.6ex\hbox{$\sim$}}}
\def\ltorder{\mathrel{\raise.3ex\hbox{$<$}\mkern-14mu
    \lower0.6ex\hbox{$\sim$}}}

\title[Long-Term Evolution of a Direct Collapse to Supermassive Black Hole Seeds]
{Supermassive Black Hole Seed Formation at High\\ Redshifts: Long-Term Evolution of the Direct Collapse}

\author[Isaac Shlosman, Jun-Hwan Choi, Mitchell Begelman and Kentaro Nagamine]
{Isaac Shlosman$^{1,2}$\thanks{E-mail: shlosman@pa.uky.edu},
Jun-Hwan Choi$^{3}$\thanks{E-mail: jhchoi@astro.as.utexas.edu}, 
     Mitchell C. Begelman$^{4,5}$, Kentaro Nagamine$^{2,6}$
\\
$^{1}$ Department of Physics \& Astronomy, University of Kentucky, Lexington, KY 40506-0055,
USA\\
$^{2}$ Theoretical Astrophysics, Department of Earth \& Space Science, Graduate School of Science, Osaka University,
     Osaka 560-0043, Japan\\  
$^{3}$ Department of Astronomy, University of Texas, Austin,  TX 78712-1205 , USA \\
$^{4}$ JILA, University of Colorado and National Institute of Standards and Technology, 
     440 UCB, Boulder, CO 80309-0440, USA\\
$^{5}$ Department of Astrophysical and Planetary Sciences, 391 UCB,  
     Boulder, CO 80309-0391, USA\\ 
$^{6}$ Department of Physics \& Astronomy, University of Nevada, Las Vegas, NV 89154-4002, USA\\
}

\begin{document}

\date{Accepted 2015 November 16; Received 2015 November 10; in original form 2015 August 18}


\maketitle

\begin{abstract}
We use cosmological adaptive mesh refinement (AMR) code Enzo zoom-in simulations to study
the long term evolution of the
collapsing gas within dark matter halos at $z$. This direct collapse
process is a leading candidate for rapid formation of supermassive black hole (SMBH) 
seeds. To circumvent the Courant condition at small radii, we apply the sink particle
method, focusing on evolution on scales $\sim 0.01-10$\,pc. The collapse
proceeds in two stages, with the secondary runaway
happening within the central 10\,pc. The sink particles form when
the collapsing gas requires additional refinement of the grid size at the highest refinement
level. Their growth is negligible
with the sole exception of the central seed which
grows dramatically to $M_{\rm seed}\sim 2\times 10^6\,M_\odot$ in $\sim 2$\,Myr,
confirming the feasibility of this path to the SMBH.
The variability of angular momentum in the accreted gas results in the formation of
two misaligned disks. Both disks lie within the Roche limit of the central seed. While
the inner disk is geometrically thin and weakly asymmetric, the outer
disk flares due to turbulent motions as a result of the massive inflow along a pair of
penetrating filaments. The filamentary inflow determines the dominant 
Fourier modes in this disk --- these modes have a non-self-gravitational
origin. We do not confirm that
$m=1$ is a dominant mode that drives the inflow in the presence of a central massive object.
The overall configuration appears to be generic, and is expected to form when the
central seed becomes sufficiently massive.
\end{abstract}

\begin{keywords}
methods: numerical --- galaxies: formation --- galaxies: high-redshift --- cosmology: theory
--- cosmology: dark ages, reionization, first stars
\end{keywords}

\section{Introduction}
\label{sec:intro}

Supermassive black holes (SMBHs) formed early in the
history of the universe. Luminous quasars have been observed
up to redshifts $z\sim 7.1$ \citep[e.g.,][]{Fan.etal:03,Mortlock.etal:11,Wu.etal:15}, 
and have been estimated to
host SMBHs in excess of $10^9\,M_\odot$, just 700\,Myr after the
Big Bang. Unless the SMBHs are primordial, they must have formed by the 
accumulation of matter during the epoch of galaxy formation --- either as 
remnants of the Population\,III stars \citep[e.g.,][]{Haiman.Loeb:01,
Abel.etal:02,Bromm.Larson:04,Volonteri.Rees:06,Li.etal:07,Pelupessy.etal:07}, or 
the end products of gas collapse into dark matter (DM) halos with
virial temperatures $\gtorder 10^4$\,K \citep[e.g.,][]{Haehnelt.Rees:93,
Oh.Haiman:02,Bromm.Loeb:03,Volonteri.Rees:06,Begelman.etal:06,Begelman.Shlosman:09,
Milosavljevic.etal:09,Mayer.etal:10,Schleicher.etal:10,Hosokawa.etal:11,Johnson.etal:11, 
Choi.etal:13,Choi.etal:15,Latif.etal:13,Prieto.etal:13}. Accounting
for radiative feedback during formation of Pop\,III
stars lowers their main sequence masses to be more in line
with those of normal stars. The black
hole remnants are downsized as well, to $\sim 10\,M_\odot$, making it 
exceedingly difficult to explain the growth of early SMBHs from Pop\,III
seeds. Furthermore,
recent results from Planck have pushed the beginning
of the Pop\,III epoch forward from $\sim 400$\,Myr to $\sim 560$\,Myr after the
Big Bang \citep[e.g.,][]{PlankCollaboration.etal:15}. Taken together, these arguments 
Favor substantially
more massive SMBH seeds, $\sim 10^5 - 10^7\,M_\odot$, as the direct
collapse models suggest.

Important details of direct collapse remain unclear,
e.g., how does the process evolve when the flow becomes 
optically thick to internally produced radiation? How is 
angular momentum redistributed during the optically-thick
regime? Does the collapse lead to a hydrostatically supported
supermassive star (SMS) --- a precursor to 
SMBH seed formation \citep[e.g.,][]{Begelman:10} --- or does the flow
remain disky and by-pass the SMS and the associated
thermonuclear stage \citep{Choi.etal:13}?

On the other hand, we now understand some of the details
of the collapse in the optically-thin regime. The fragmentation
of the flow is damped by the dominant gravitational potential of
the host DM halo and by virial supersonic turbulence ---
both of which substantially increase the Jeans mass of the gas.
Furthermore, the DM acts as a sink of the angular momentum from the 
collapsing gas, both
in isolated collapse models and in the cosmological context 
\citep[e.g.,][]{Wise.etal:08,Begelman.Shlosman:09,Choi.etal:13,Choi.etal:15}. 

Studies of the direct collapse to form the SMBH seeds
involve computationally-intensive efforts to follow the hydrodynamics
and the thermal state of the collapsing gas.
Processes like fragmentation, star formation, redistribution
of angular momentum in the collapsing flow, development
of a supersonic turbulence, etc., can be addressed only numerically.

Typically, the collapse region extends over many orders
of magnitude in radius, from a few kpc down to $\sim 1-100$\,AU.
Even sophisticated methods face difficulties in dealing with such an
extended dynamic range. As the collapse progresses, the
simulation timesteps become increasingly short, making it prohibitively 
expensive to follow the evolution. Moreover,
the buildup of the optical depth in the flow can fully or
partially trap the radiation produced in situ, affecting the
dynamics and thermodynamics of the collapsing gas.

Two options have been proposed to overcome these obstacles.
The first involves (consecutive) zoom-in simulations
typically associated with calculations of cosmological structure
formation and galaxy evolution 
\citep[e.g.,][]{Becerra.etal:15}. This method allows one to increase
the mass and spatial resolution of the model, especially on the
smallest scales, to introduce various physical processes, on different
scales, and to avoid artefacts associated with a small number of particles or
grid cells at small radii. \citet[][]{Becerra.etal:15}
have assumed that the radiative cooling becomes inefficient at
some presumed radius. Such an exponential cutoff in the cooling rate
naturally leads to a radiation pressure-dominated entity --- an 
SMS. By itself this does not resolve the
issue of radiation trapping in the flow --- in a 3-D rotation-dominated
inflow the radiation, in principle, can escape.
It does not allow one to study the optically thin-to-thick transition in the
flow and the nature of the flow inside the optically-thick region.
Merely increasing the mass and spatial resolutions will not resolve 
the problem and will not allow one to prolong the simulation without also 
introducing additional physical processes such as radiative transfer and
associated dynamical effects. At present time, sufficiently powerful radiation
hydrodynamics codes are not generally available.

The second option involves the sink particle method,
successfully applied to star formation and other problems
to focus on
a specific range in radius, preventing the smaller scales from
forcing an increasingly small timestep, and thus enabling one to follow
the long-term evolution of the collapsing flow \citep[e.g.,][]{Bate.etal:95,
Krumholz.etal:04,Federrath.etal:10,Wang.etal:10,Teyssier.etal:11,Gong.Ostriker:13}. 
In this method, gas on small spatial scales
is replaced by the sink particles, which can grow in mass as
the surrounding gas evolves. 

In this paper, we apply the sink particle
method to the direct collapse problem in a cosmological framework. We focus 
on the long-term evolution of the flow and on associated
processes on scales of $\sim 0.01-10$\,pc, deep inside the DM halo 
gravitational potential. The choice of this scale is based on purely numerical 
reasons, in order to circumvent the exceedingly small timestep. It also allows us
to avoid additional physical processes on the small scale, such as radiative
transfer, without any loss of generality.

Starting with our cosmological
simulations of direct collapse \citep{Choi.etal:15}, we create
sink particles in the flow (as described in Section\,\ref{sec:method}), and analyze
motions and growth of these particles and their effect
on the collapse itself. Choi et al. has followed the collapse
down to $\sim 10^{-4}$\,pc scales, where one anticipates trapping
of the continuum photons produced in the flow and the formation
of a photosphere. Of course, the Lyman$\,\alpha$ photons are trapped
or at least partially trapped already at larger radii.
Here we ignore scales smaller than $\sim 0.01$\,pc, which allows us 
to continue following the collapse on scales  $> 0.01$\,pc, over
timescales much longer than those simulated by Choi et al.

This paper is structured as follows. Section\,\ref{sec:method} provides
the details of our numerical technique and explains the sink
particle method implemented here. We present
our results in Section\,\ref{sec:sink} and discuss them in the last section.

\section{Numerical techniques}
\label{sec:method}

\subsection{Simulations}
\label{sec:sims}

We use the Eulerian adaptive mesh refinement (AMR) code
Enzo-2.3, which has been tested extensively and is publicly
available \citep{Bryan.Norman:97,Norman.Bryan:99,
Bryan.etal:14}. Enzo uses a particle-mesh $N$-body
method to calculate the gravitational dynamics, including
collisionless DM particles, and a second-order piecewise
parabolic method \citep[PPM,][]{Bryan.etal:95} to solve hydrodynamics.
The structured AMR used in Enzo places no
fundamental restrictions on the number of rectangular grids
used to cover some region of space at a given level of refinement,
or on the number of levels of refinement \citep{Berger.Colella:89}. A 
region of the simulation grid is refined by a
factor of 2 in lengthscale, if either the gas or DM density become
greater than $\rho_{\rm 0,gas,dm}N^l$, where $\rho_{\rm 0,gas,dm}$ is the 
cosmic
mean density for the gas or DM respectively, $N = 2$ is the
refinement factor, and $l$ is the maximal AMR refinement
level. 

The Jeans length has been resolved by at least 16 cells in the simulations.
Hence, the \citet{Truelove.etal:97} requirement for resolution of the Jeans 
length (see Section\,2.3), i.e., at least four cells, has been superseded 
substantially \citep[e.g.,][]{Sur.etal:10,Federrath.etal:11,Turk.etal:12,
Latif.etal:13}.

Enzo follows the non-equilibrium evolution of six species:
H, H$^+$, He, He$^+$, He$^{++}$, and $e^-$ \citep{Abel.etal:97,
Anninos.etal:97} in a gas with a primordial composition. It calculates
radiative heating and cooling following atomic line excitation,
recombination, collisional excitation and free-free
transitions. Radiative losses from atomic cooling are computed
in the optically-thin limit.

Several radiation processes
which prevent H$_2$ formation in the collapsing flow have
been proposed \citep[e.g.,][]{Omukai:01,Spaans.Silk:06,Schleicher.etal:10,
Latif.etal:11,Choi.etal:13,Inayoshi.etal:14,Sugimura.etal:14}. In
this work we assume that H$_2$ does not form and exclude the
chemistry and cooling related to H$_2$, which simplifies the
chemical evolution.

\subsection{Zoom-in initial conditions}
\label{sec:ICs}

We follow the long-term dynamical evolution of the collapsing
gas within a DM halo in a fully cosmological environment
and subject to atomic cooling. To satisfy the resolution
requirement, we use the MUSIC code \citep{Hahn.Abel:11}
to generate the cosmological zoom-in initial conditions
(ICs). MUSIC uses a real-space convolution approach
in conjunction with an adaptive multi-grid Poisson solver to
generate highly accurate nested density, particle displacement,
and velocity fields suitable for multi-scale zoom-in
simulations of structure formation in the universe. 

Generating
a set of ``zoom-in'' ICs is a two-step process. First,
we generate $1h^{-1}$\,Mpc comoving $128^3$ DM-only ICs for the
pathfinder simulation and run it without AMR until $z = 10$.
Using the HOP group finder \citep{Eisenstein.Hut:98}, we select
an appropriate DM halo, whose mass is $\sim 10^8 
h^{-1}\,{\rm M_\odot}$
at $z = 10$. Secondly, we generate $1h^{-1}\,{\rm Mpc}$ ICs with 
$512^3$ resolution
in DM and gas in the zoom-in region. Since we use the same random seeds
of these ICs as the ICs at the first step, the phases of both
ICs are identical. The zoom-in region is centered on the selected
halo position and is set to be large enough to cover
the initial positions of all selected halo particles. 

The ICs are generated using WMAP5 cosmology:
$\Omega_{\rm \Lambda} = 0.721$, $\Omega_{\rm m} = 0.279$, $\Omega_{\rm b}
= 0.0445$, $h=0.701$, $\sigma = 0.807$, and $n_{\rm e} = 0.961$.
In the following we use $R$ for
spherical coordinates and $r$ for cylindrical ones.

\subsection{Sink particle algorithm}
\label{sec:sink_intro}

Sink particles are used in order to weaken the Courant conditions
restrictions on the timestep. The restriction can result
from the increase in the local density of the gas and demand for
additional refinement (i.e. spatial resolution).
Under these circumstances, it is nearly impossible to study the large-scale, 
long-term evolution of the direct collapse within DM halos.
To circumvent these numerical obstacles, we adopt the sink particle
method to model the gravitational collapse.
This is especially
relevant for the present work, which aims at long-term evolution
of the direct collapse. Before the gas density exceeds
that of the DM density, at $t\sim 350$\,Myr, sink particles are not used.
However, after the gas decouples, its central density
increases rapidly.
Below we describe the specifics of this method.

Various works have invoked sink particles in order to advance their goals
\citep[e.g.,][]{Bate.etal:95,Krumholz.etal:04,Federrath.etal:10,Wang.etal:10,
Teyssier.etal:11,Gong.Ostriker:13}.
We implement the sink particle method largely based on \citet{Wang.etal:10}, 
who studied massive star formation using Enzo AMR simulations.
We require a sink particle to form when the cell violates 
the refinement criterion at the highest refinement level in the collapsing part of 
the flow. 

When this happens,
a sink particle is inserted at the center of the cell.
The initial sink particle mass is computed based on the mass exceeding 
the maximum allowed density at a maximum refinement level.
In other words, each cell density has a maximal value
that does not violate the refinement criterion.
The initial sink particle velocity is calculated based on linear momentum 
conservation.

After a sink particle is formed, it accretes the gas from its
host cell according to following prescription:
\begin{eqnarray}
\dot{M}_{\rm sp} = 4 \pi \rho_{\rm out} r_{\rm B}^2 \sqrt{1.2544 c_{\rm cell}^2 +
   v_{\rm rel}^2},
\end{eqnarray}
where $\dot{M}_{\rm sp}$ is the mass growth rate of a sink particle, 
$\rho_{\rm out} =
\rho_{\rm cell} \, {\rm min}[1.0, (l_{\rm cell}/r_{\rm B})^{1.5}]$, $l_{\rm cell}$
is the the cell size, $r_{\rm B} = GM_{\rm sp}/(c_{\rm cell}^2 + v_{\rm rel}^2)$
is the Bondi accretion radius, $c_{\rm cell}$ is the sound speed in the cell, and
$v_{\rm rel}$ is the relative velocity between the host cell and the sink
particle.

Merging between sink particles is a necessary process in order to accurately
estimate their mass growth rate and to reduce the computational cost.
In principle, the sink particles represent overdense gas clumps in our
simulations whose internal evolution we ignore. 
We require two sink particles to merge when their separation becomes smaller than
5$l_{\rm cell}$, and assume that the less massive sink particle merges with the 
more massive one. The merging process is quite insensitive to the
definition of the critical separation between particles 
\citep[e.g.,][]{Wang.etal:10}. We have confirmed this by comparing various maximal 
refinement levels.

In this paper, we use three different maximum refinement levels that affect the overall 
simulation resolution and the scale of the sink particle formation. The maximal
AMR refinement level $l$ (see Section\,\ref{sec:sims}) is 10 in run L10, $l=12$ in run L12, 
and $l=15$ in run L15. The maximal spatial resolution 
corresponds to $\sim 0.1$\,pc for run L12 and to $\sim 0.01$\,pc for run L15.

All three simulations have been restarted from the cosmological
run described in \citet{Choi.etal:15}, at $t\sim 355$\,Myr,
when the refinement level in the gas is $l = 8$, and when the gas-to-DM 
density ratio in the central $\sim 10$\,pc is about unity. They are identical
simulations except for the maximal AMR level. While our analysis here is based on
run L12, we use runs L10 and L15 to test convergence
of the observed evolution.

\section{Results}
\label{sec:sink}

The DM halo under investigation has been selected from
a computational box containing several such objects at
$z = 10$. This halo has been re-simulated at high resolution
down to $z\sim 12$, or $t\sim 355$\,Myr from the Big Bang, when it
reached the virial mass and radius of $M_{\rm h}\sim 2\times 10^7h^{-1}\,M_\odot$
and $R_{\rm vir}\sim 10h^{-1}$\,kpc in comoving coordinates, with a cosmological
spin parameter $\lambda\sim 0.03$. At this time, the DM density profile
can be approximated by an NFW profile \citep[e.g.,][]{Navarro.etal:97}, with
a characteristic scale radius of $R_{\rm s}\sim 30$\,pc,
in physical coordinates, and with a concentration parameter 
$c\equiv R_{\rm vir}/R_{\rm s}\sim 25$. At about $t\sim 355$\,Myr, the gas-to-DM
density ratio exceeds unity interior to $\sim R_{\rm s}$. The gas density
quickly establishes a density profile slightly steeper
than $\sim R^{-2}$, down to the limiting radius of $R\sim 10^{-4}$\,pc,
imposing a tight Courant condition on the timestep.

As shown by \citet{Choi.etal:13,Choi.etal:15}, direct collapse, in both the isolated
case and in a cosmological framework, can be divided into
two parts: the first stage ends when the central gas density reaches that of the
DM, and the second stage begins when the central gas density exceeds that of the DM, and
the gas effectively decouples from the DM. During this latter stage, 
an $r^{-2}$ density profile is established all the way down to $\sim 10^{-4}$\,pc 
--- the region
where the optical depth to continuum radiation produced by the accretion flow becomes
larger than unity, the so-called thin-to-thick transition. As such a large dynamic range
imposes a very strict condition on the timestep, and because we focus on the optically-thin 
part of the flow, we introduce sink particles based on the criteria discussed
in Section\,2.3. By doing this, we remove the smallest spatial scales, $r < 0.1$\,pc
(for run L12) and  $< 0.01$\,pc (for run L15), from consideration. Hence, these are  
the limiting resolutions of the models.
On the other hand, we gain the ability to follow the collapse well beyond the time
achieved in \citet{Choi.etal:15}.

\subsection{Rapid growth of the central sink particle}
\label{sec:growth}

\begin{figure}
\centerline{
  \includegraphics[width=0.57\textwidth,angle=0] {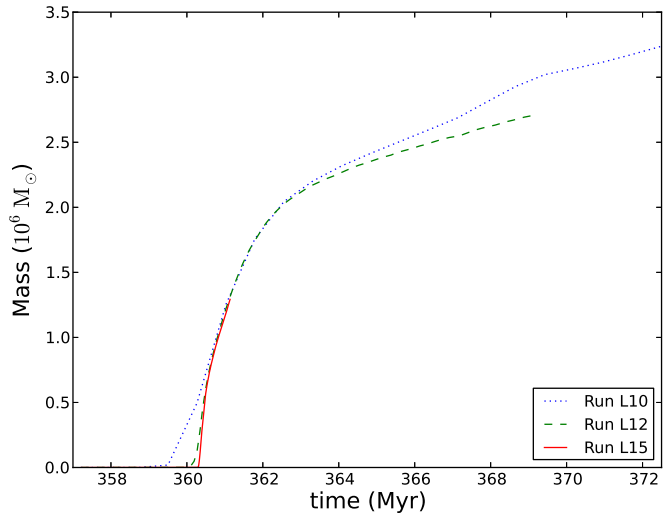}
}
\caption{
Evolution of the central sink particle mass, $M_{\rm seed}$, forming at the position 
of the density peak for three different AMR runs, L10, L12 and L15.
While L10 exhibits a somewhat slower or faster growth of the central seed
at various times, L12 and L15 show virtually no difference.
Note, that the initial growth of the seed mass is very rapid, so that it reaches 
$M_{\rm seed}\sim 2\times 10^6\,M_\odot$ in $\ltorder 2$\,Myr.
}
\label{fig:SinkEvolution}
\end{figure}
\begin{figure*}
\centerline{
   \includegraphics[width=1.10\textwidth,angle=0] {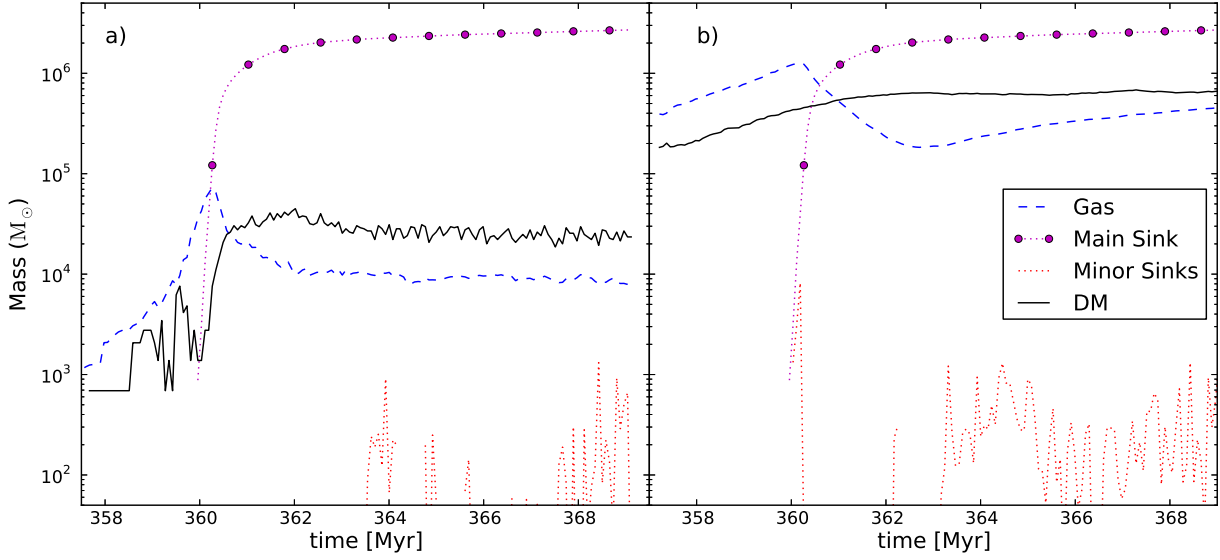} 
}
\caption{Evolution of masses within spherical volumes with a radius of 2\,pc (left frame) and
20\,pc (right frame). Shown are the gas mass (blue dashed line), DM mass (black solid line),
central seed (dotted line with large dots), and the total of all minor sink particles 
(dotted green line).
}
\label{fig:masses}
\end{figure*}

The masses of sink particles introduced in the run L12 
correspond to the gas mass of the smallest grid cells. Runs
L10 and L15, in this respect, are very similar, but their resolutions are $\sim 1$\,pc and
$\sim 0.01$\,pc respectively. So, the newly born particles 
provide only a negligible effect on the flow when they are introduced.
They can grow by  accretion and merging and, at least in principle, their
influence grows. We find that sink particles appear always within the central $\sim 1-2$\,pc,
measured from the densest gas cell. Their number never substantially exceeds $\sim 10$
in the collapsing flow. They grow, but 
their individual masses stay below $\sim 10^3\,M_\odot$.
The sole exception is the sink particle that lies at the center. This particle
grows fast, and in less than $\sim 2$\,Myrs surpasses $M_{\rm seed}\sim 2\times 10^6\,M_\odot$, as shown in
Figure\,\ref{fig:SinkEvolution}. (We term this sink particle as a `seed' and conjecture that its evolution is
closely related to the future formation of the SMBH seed.) After this time, its growth rate 
saturates substantially. We separate these phases
into dynamic and secular growth stages of the central seed. 

For comparison, we also show the evolution of the central sink particle in runs L10 and L15.  
In run L10, the central sink particle starts to grow slightly earlier, but very 
quickly its
growth curve merges with that of L12, even before the end of the dynamic stage. During
the secular stage, the L10 growth rate initially exceeds that for L12.
Asymptotically, the L10 and L12 growth rates are nearly identical, but the final seed
masses differ by about 20\%.

Run L15 displays a nearly identical growth for the central particle, except for very early
times --- growth is triggered slightly later but the rate is higher. We therefore, 
conclude
that the resolution of run L12 is sufficient for our purpose, because the growth rates in 
L12 and L15 appear identical.

So, irrespective of the resolution we choose for the sink particles, the central seed grows
in a very short time to exceed $10^6\,M_\odot$. How is this mass converted into the SMBH seed
is a subject for future research.

\subsection{The central massive seed and disk formation}
\label{sec:disk}

To understand the specifics of the growth of the central seed, we plot the evolution of the
gas masses within the central 2\,pc and 20\,pc, and compare them with the mass of the central
seed, the total mass of other (off-center) sink particles, and the DM within these spherical 
volumes.
Note that the sink particles grow only by accreting gas and incorporating other sink
particles, but not through accretion of DM. The total baryonic mass, i.e., the gas and the sink particles,
is, therefore, conserved.

In Figure\,\ref{fig:masses}a, we observe that the total mass of the off-center sink particles in the region
remains very small. The gas accumulation inside the innermost 2\,pc starts
around 358\,Myr and peaks at $\sim 360$\,Myr, prior to and concurrently with the formation 
and rapid growth
of the central seed. After the rapid growth stage, the gas mass inside this region is
remarkably constant at $\sim 10^4\,M_\odot$, showing that {\it net} inflow into this
region is negligible. Below, we discuss the detailed mass balance in the region. Note, that
the gas at the peak inflow rate into this region drags in
the DM, which dominates the gas mass during the secular evolution of the central sink.
This provides additional stability against the gas fragmentation in the region.

\begin{figure}
\centerline{
  \includegraphics[width=0.55\textwidth,angle=0] {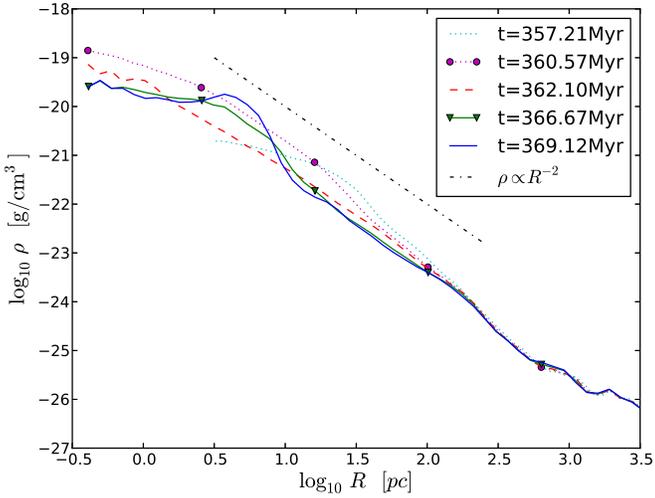}
}
\caption{Evolution of the gas density 
profile for run L12. The density profiles, $\rho\sim r^{-\alpha}$, exhibit flattening in the 
central region, from $\alpha\sim 2$ to $\alpha\sim 0.5$, where the gas assembles into a disk 
structure around the central massive seed.
}
\label{fig:profiles}
\end{figure}
\begin{figure}
\centerline{
  \includegraphics[width=0.55\textwidth,angle=0] {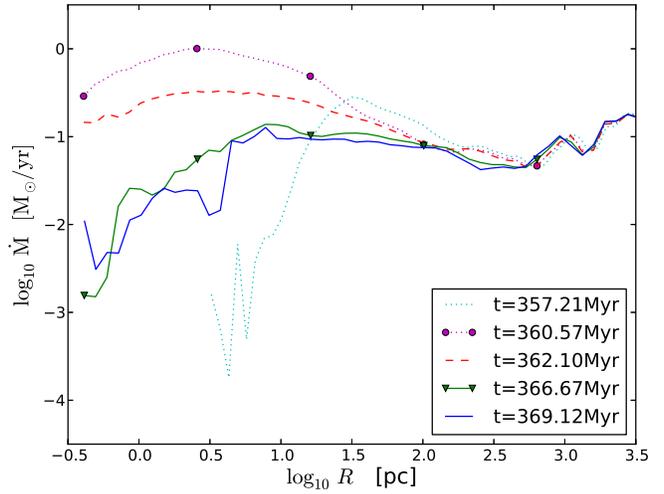}
}
\caption{
Evolution of the mass accretion rate profile for the L12 run. Note the sharp decrease in 
$\dot M$ from
the peak of $\sim 0.3-1\,M_\odot\,{\rm yr^{-1}}$ by about two orders of magnitude
at smaller radii, as a result of the rotational support achieved near the massive
sink particle.
}
\label{fig:Mdot}
\end{figure}
\begin{figure}
\centerline{
  \includegraphics[width=0.56\textwidth,angle=0] {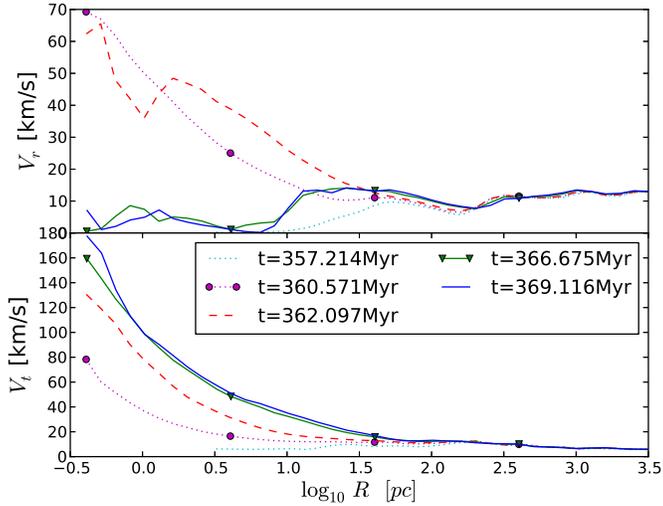}
}
\caption{
Evolution of the radial (top) and tangential (bottom) velocity profiles, $v_{\rm r}$ and
$v_{\rm t}$, in the gas for run L12. Note the sharp decrease in the central $v_r$
and increase in the $v_{\rm t}$ there --- a clear sign of disk formation.
}
\label{fig:vprof}
\end{figure}

\begin{figure}
\centerline{
  \includegraphics[width=0.57\textwidth,angle=0] {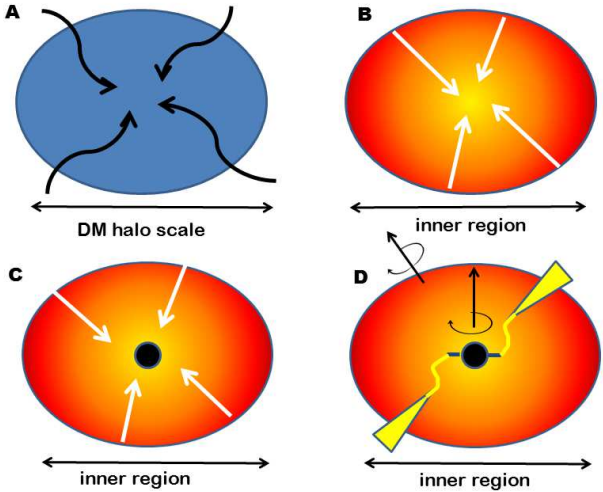}
}
\caption{
Schematic picture of various stages in the direct collapse with the rapidly growing central
seed. {\it Stage A:} Filamentary inflow into the DM halo on scale of $\sim 1$\,kpc; {\it Stage B:} 
Gas density exceeds the
DM density and the secondary collapse is triggered. In the case of current simulation setup, the 
`inner region' indicates roughly 20\,pc region from the sink; {\it Stage C:} Central sink forms and rapidly 
grows to $\sim 10^6\,M_\odot$; {\it Stage D:} High angular momentum gas accumulates in a system of
misaligned disks (shown here edge-on), forming due to the fluctuating 
angular momentum of the collapsing gas. The small inner disk is shown horizontal,
and the outer disk is inclined. The disks are connected by a symmetric warp. See also 
Figure\,\ref{fig:VelDen} and explanations in the text.
}
\label{fig:draw}
\end{figure}

The mass evolution within the central 20\,pc paints a similar picture, with an important
difference (Fig.\,\ref{fig:masses}b).
The gas mass grows monotonically until the formation of the central seed, then drops by
an order of magnitude during the 2\,Myr rapid growth of the central seed. This
explains the source of the mass in this seed. After this time, the gas mass continues to
accumulate at nearly the original rate. The DM here experiences the same adiabatic contraction
as in Figure\,\ref{fig:masses}a, and dominates the gas. It is clear, however, that the gas will 
shortly reverse this situation --- as shown by Figure\,\ref{fig:masses}b, about 10\,Myrs after the 
formation of the 
central seed, the gas mass in the region will surpass that of the DM, yet it will remain
well below the mass of the central seed.
We discuss reasons for the gas evolution in the vicinity of the central seed and its global consequences
below and in Section\,\ref{sec:discuss}.

The differences between gas evolution inside the central region and on larger scales 
is related to the gas kinematics. This is seen in Figure\,\ref{fig:profiles},
where the slope of the gas density within the central $\sim 10$\,pc, $\rho\sim r^{-\alpha}$,
changes from $\alpha\sim 2$ to $\sim 0.5$ --- the gas density profile has flattened. Because we 
assume that the gas is optically-thin, the temperature profile remains the same as shown in
\citet{Choi.etal:15}.

Another by-product of the change in the prevailing kinematics, is the evolution of the 
gas accretion rate profile.
The peak accretion rate of $\sim 1\,M_\odot\,{\rm yr^{-1}}$ inside $\sim 10$\,pc 
is reached between $t\sim 360-362$\,Myr (Fig.\,\ref{fig:Mdot}). This is reflected in the very fast growth 
of the central seed during these 2\,Myr.
Immediately after this, the radial profile of the mass accretion rate, $\dot M(r)$,
drops dramatically from its peak by about two orders of magnitude or more with decreasing radius. That
is, at $r\sim 10$\,pc, the mass inflow rate stays approximately the same, 
$\sim 0.3\,M_\odot\,{\rm yr^{-1}}$, but it declines sharply with radius to 
$\sim 3\times 10^{-3}\,M_\odot\,{\rm yr^{-1}}$ at $\sim 0.3$\,pc. Note, that substantial amount
of gas, $\sim 6\times 10^6\,M_\odot$, resides in the DM halo outside 20\,pc at the time of disk formation, 
and that the accretion rate onto the halo remains high as well.

This behavior in the gas density profile and its accretion rate within the central 20\,pc
means that the gas accumulates there at rapid pace. At around $t\sim 360$\,Myr, this is related
to the formation and a rapid growth of the central seed. However, after $t\sim 362$\,Myr,
this is the direct consequence of the accumulation of angular momentum and formation of a 
gaseous disk in this region.

\begin{figure*}
\includegraphics[width=1.02\textwidth,angle=0] {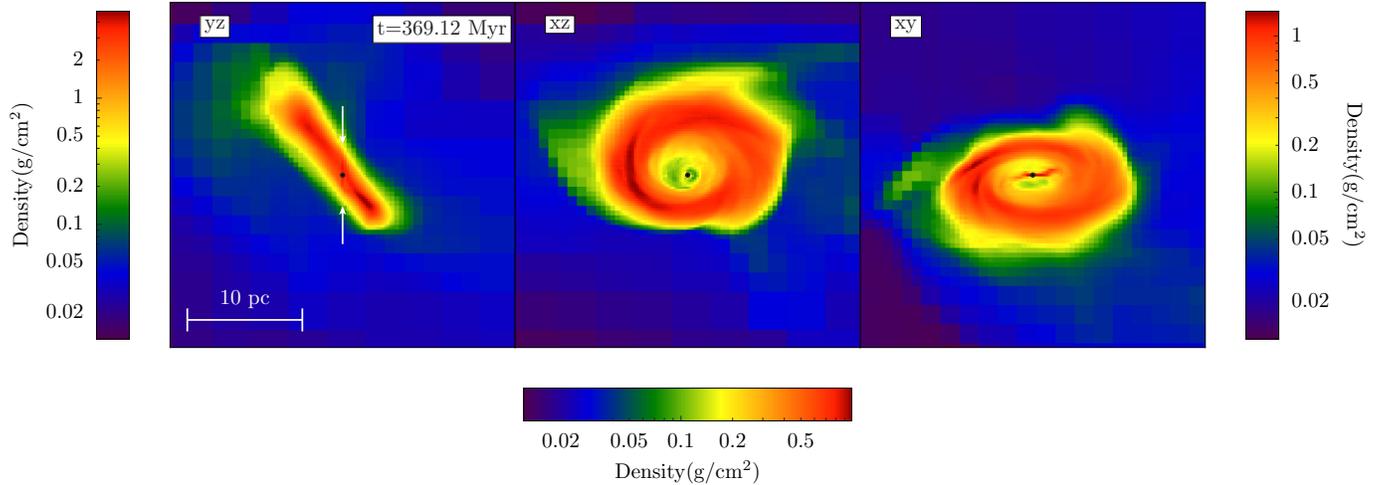}
  \caption{
Gas column density normal to main planes at the end of the simulation, $t\sim 369$\,Myr,
for run
L12. The left frame shows the inner edge-on disk (position delineated by the white arrows)
embedded in the outer disk
which is inclined by about $45^\circ$ in the $yz$ plane. The middle frame shows the same
figure with the inner disk being face-on in the $xz$ plane. The right frame shows another
projection of the inner edge-on disk and the inclined (toward the observer) outer disk in the
$xy$ plane. The inner warp connection between the disks can be clearly seen in this projection.
The position of the central seed is shown by a black point. The filaments which feed the outer
disk are clearly visible.
  }
\label{fig:Project}
\end{figure*}

Evolution of the mass accretion rate is clearly reflected in the abrupt collapse
of the gas and formation of the central seed, and can be followed in the top frame of 
Figure\,\ref{fig:vprof}. The radial
velocity at $t\sim 360.6$\,Myr increases to $\gtorder 70\,{\rm km\,s^{-1}}$. This is well
above the virial velocity of the DM halo, by almost an order of magnitude, and can only result 
from the gravitational decoupling of the gas from the background DM potential.
Rotational velocity around 0.3\,pc from the central seed has reached $v_{\rm t}\sim
160\,{\rm km\,s^{-1}}$ by  $t\sim 363.6$\,Myr --- exceeding the DM virial velocity by a 
factor of $\sim 16$ (Fig.\,\ref{fig:vprof}, bottom frame).
Hence the depth of the potential well due to the massive seed has substantially surpassed 
that in the DM.
Basically, this corresponds to the following sequence of events: onset of
gravitational collapse in the gas on 
scales $\ltorder 20$\,pc, gravitational decoupling of the gas from the DM, formation
of the massive seed, and continuing accretion onto this seed.
At the same time, the increased rotational support for the gas can
be clearly observed in a decreased radial inflow velocity to $v_{\rm r} < 10\,{\rm km\,s^{-1}}$, 
and a corresponding sharp increase in the tangential velocity, $v_{\rm t}$, which develops 
a Keplerian profile inside the inner $r\sim 50$\,pc, as seen in Figure\,\ref{fig:vprof}.

The central seed has reached $M_{\rm seed}\sim 10^4\,M_\odot$ at $t\sim 360.1$\,Myr. Strong 
compression of the collapsing gas, shocks, and radial filaments can be observed at this
time. The disk first becomes visible at $\sim 361.5$\,Myr, when 
$M_{\rm seed}\sim 1.6\times
10^6\,M_\odot$, and can be seen in two out of three projection planes, $xz$ (face-on) and
$xy$ (edge-on), initially on scales of the inner $2-3$\,pc. It grows rapidly in size,
as a strong gaseous bar with two associated {\it open} spirals\footnote{Such spirals signal
efficient angular momentum loss by the gas.}  become dominant around $\sim
362.1$\,Myr. By $\sim 362.7$\,Myr, the disk radius has reached $\sim 5$\,pc. Around $\sim
363.7$\,Myr, one starts to distinguish between an inner, geometrically thin disk, 2-3\,pc 
in radius, and an outer, much thicker disk, which is misaligned by about $45^\circ$ with inner 
one. The overall radius of this configuration is about 10--12\,pc. Formation of the outer
inclined disk is the result of gas influx with a different direction of
angular momentum into the region.

The stages of direct collapse are depicted schematically in Figure\,\ref{fig:draw}. Stage A
of the collapse corresponds to the initial inflow into the DM halo. Stage B corresponds to the
gas density exceeding that of the DM and dynamically decoupling from the background DM
potential. Stage
C culminates with the formation of the central seed and its runaway growth, and the last
stage is associated with the formation of the misaligned disk system due to the
fluctuating angular momentum of the accreting gas.

\begin{figure*}
  \includegraphics[width=1.\textwidth,angle=0] {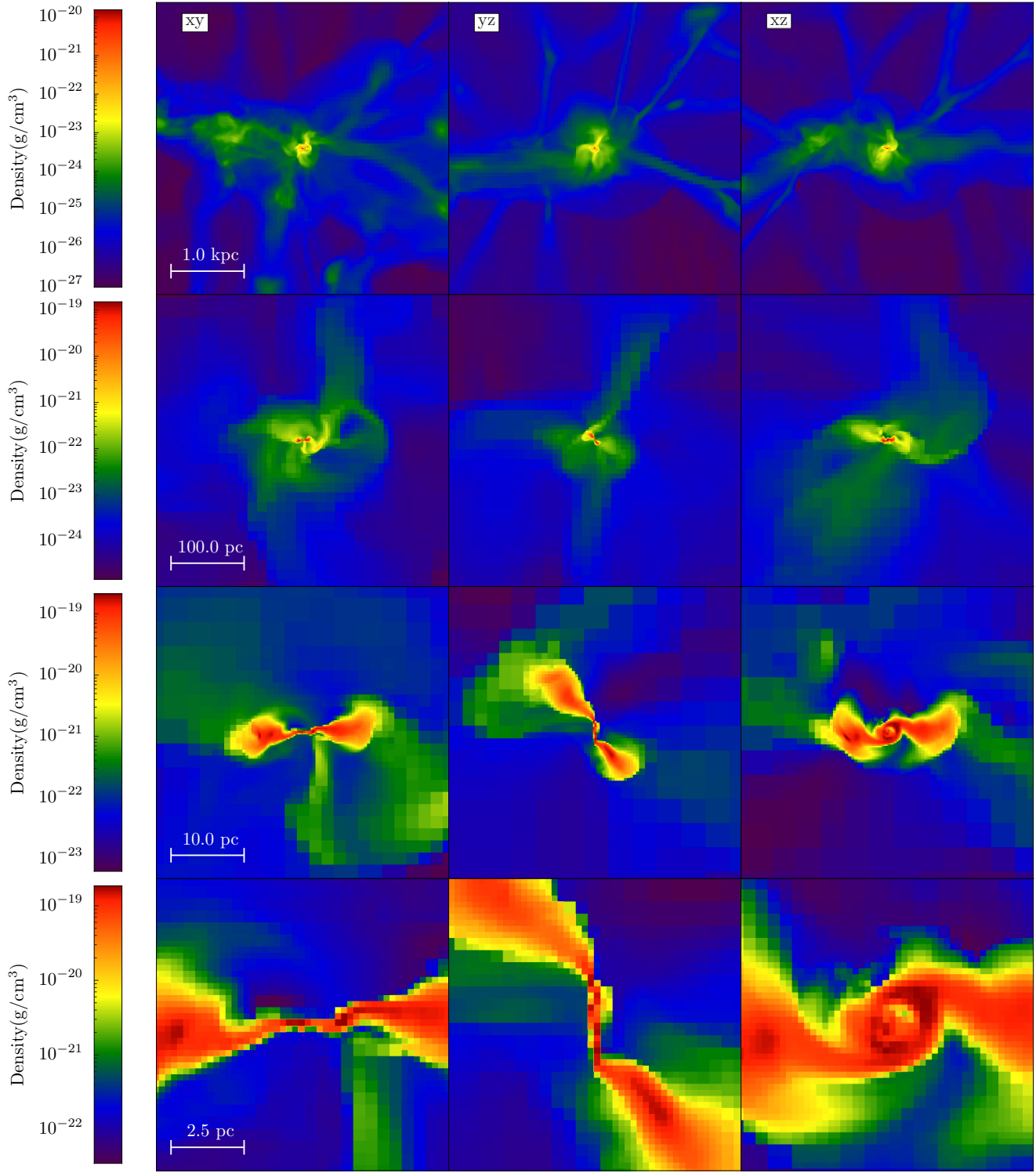}
  \caption{
Gas density slices in three projections at the end of the simulation, $t\sim 369$\,Myr, for 
run L12. The frames spanning 40\,pc and 10\,pc (on a side)
exhibit prominent disk features with the $xy$ and $yz$ planes displaying nearly edge-on 
views, and the $xz$ plane showing a nearly face-on view of the forming disk around the most
massive sink particle. 
  }
\label{fig:VelDen}
\end{figure*}

This configuration, of inner and outer disks being strongly misaligned, appears stable for the
next $\sim 6$\,Myr --- a time period which corresponds to about 20--25 rotational periods at
$\sim 15$\,pc.
By the end of the simulation, $t\sim 369.12$\,Myr, the inner disk has a radius of 3-4\,pc
and the outer disk about 10-12\,pc (Fig.\,\ref{fig:Project}). The thickness of the inner disk 
is $\ltorder 1$\,pc,
and that of the outer disk varies from $\ltorder 1$\,pc to $\sim 3$\,pc --- this disk flares 
as seen in Figure\,\ref{fig:VelDen} (bottom frames).  The inner disk is connected to the outer one
with a clearly visible symmetric, integral-shaped  warp.

We do not observe any fragmentation in the inner and outer disks.  Why is the fragmentation
suppressed in the disks, where a high-density, low-temperature gas resides? The answer to this 
question can be found in Figure\,\ref{fig:masses}a. After the formation
of the massive central seed, the gas inflow into the central 2\,pc is minimal, while $\dot M$
is high into the outer disk (see also Fig.\,\ref{fig:Mdot}).
The mass within the region, $R\ltorder 20$\,pc, is dominated by the central seed and by the 
DM that was dragged
inward by collapsing gas, in adiabatic contraction. The ratio of the gravitational
acceleration due to the gas to that of the central seed, $M_{\rm gas}(<R)/M_{\rm seed}$ is
$< 10^{-2}$ for the inner disk and $< 10^{-1}$ for the outer disk, even at the
end of the simulation. 

\begin{figure*}
\centerline{
  \includegraphics[width=1.\textwidth,angle=0] {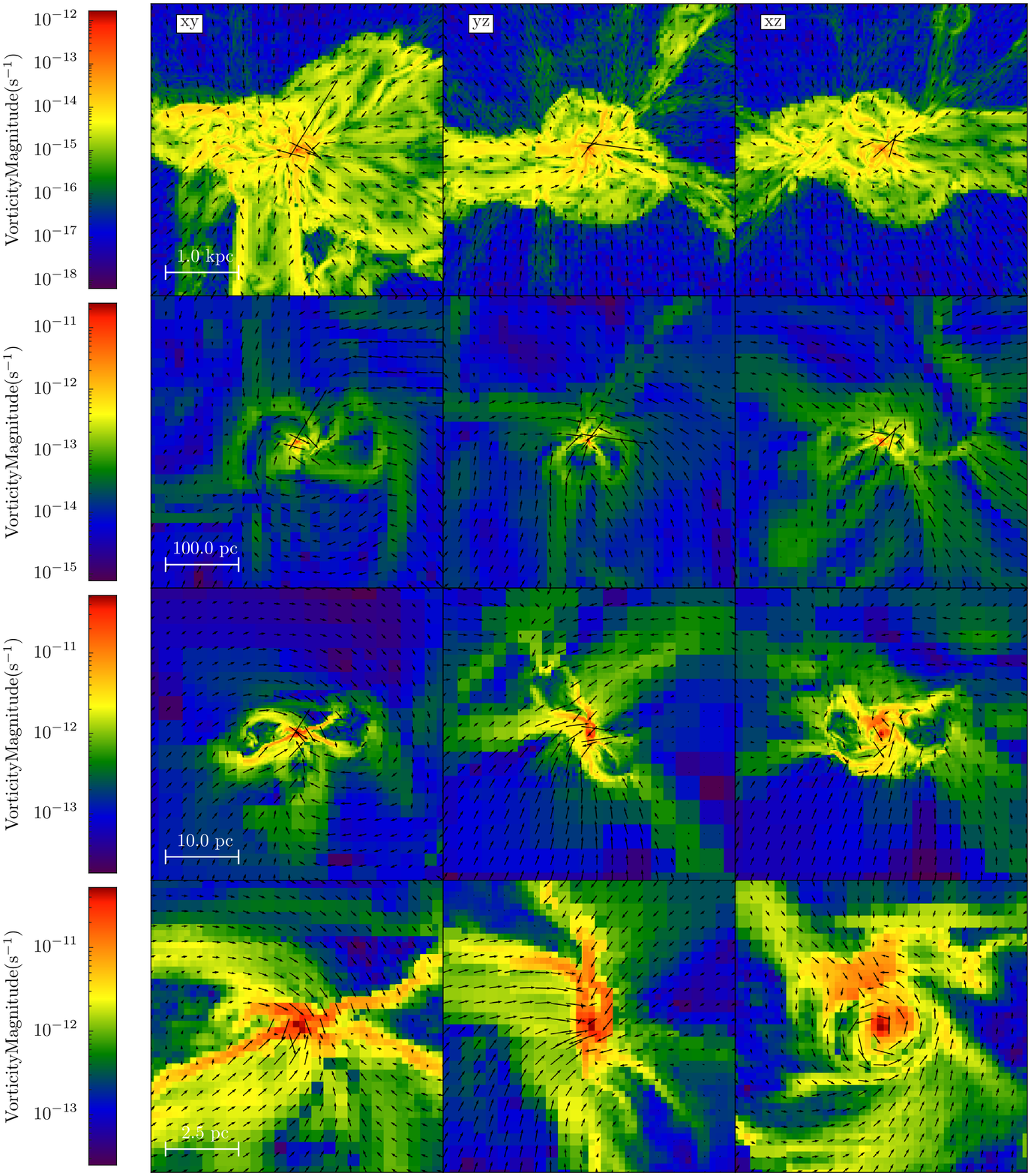}}
  \caption{
Vorticity magnitude slices of the gas evolution with projected velocity arrows, at the
end of the simulation, $t\sim 
369$\,Myr, for run L12. Velocity arrows confirm the
dominant rotation in the $xz$ plane. Arrows in the $xy$ and $yz$ planes show a shock forming at 
the disk surfaces in the gas which is accreted from outside the disk plane. 
  }
\label{fig:Vorticity}
\end{figure*}

The necessary condition for the gas to collapse in the presence of a tidal field is that its density 
has to exceed the mean density associated with the tidal field. In our case this leads to the
Roche limit radius of $\sim 10$\,pc.
Moreover, Toomre's parameter $Q\gg 1$ in both
disks, due to the presence of the central seed, and the cooling floor temperature of atomic gas.
The penetrating filaments also inject gas and stir turbulent 
motions in the outer disk (Fig.\,\ref{fig:Vorticity}). So, at least for $\sim 10$\,Myr
after the formation of the seed, and until the gas builds up anew in the region (Fig.\,\ref{fig:masses}b),
the fragmentation in the disks will be heavily suppressed.

To visualize the morphology of the gas inflow, we produce density slices in three
projection planes, $xy$, $yz$ and $xz$, using nested boxes, from 4\,kpc on a side to
10\,pc, at $t\sim 369$\,Myr (Fig.\,\ref{fig:VelDen}). The inner and outer disks are visible in the 
last two frames, i.e., 40\,pc
and 10\,pc. The inner disk is geometrically-thin and seen edge-on in the $xy$ and $yz$ planes 
and face-on in the $xz$ plane. It is surrounded by the geometrically-thicker disk,  
whose mid-plane is inclined to that
of the inner disk by about $45^\circ$ in the $yz$ plane. Spiral structure is clearly observed 
in both disks.
The accreting gas is shocked at the disk surface, as can be seen in the bottom frames
of Figure\,\ref{fig:Vorticity} (left and middle columns). Similar behavior has been observed 
by \citet{Choi.etal:13} for isolated models of direct collapse.


\subsection{Evolution of angular momentum in the accretion flow: understanding the 
dynamical consequences of the massive seed}
\label{sec:ang_mom}

\begin{figure}
\centerline{
 \includegraphics[width=0.55\textwidth,angle=0] {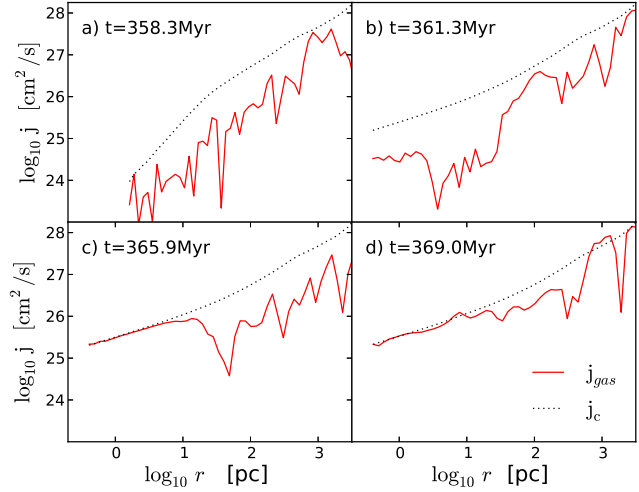} 
}
\caption{
Evolution of the specific angular momentum profile in the gas, $j(r)$,
for run L12 at 358.3\,Myr (top left), 361.33\,Myr (top right), 365.9\,Myr
(bottom left), and 369.0\,Myr (bottom right). The angular momentum is measured in a cylindrical
volume with a thickness of 10\,pc, with respect to the rotation axis of the inner disk, which happens to
nearly coincide with the $y$-axis of the computational box. The dotted lines 
represent the instantaneous Keplerian specific angular momentum
profiles. Note that the innermost $j_{\rm gas}$ reaches its Keplerian value at later times.
}
\label{fig:jyevol}
\end{figure}

An alternative way to view the growing disk in the center is to study the
evolution of the angular momentum in the collapsing flow (Fig.\,\ref{fig:jyevol}). We show the specific
angular momentum profile, $j_{\rm gas}(r)$, at four representative times. For comparison, 
we also plot the circular angular momentum profiles, $j_{\rm c}(r)$. Initially, the gas
has $j_{\rm gas} \ll j_{\rm c}$, but increases its specific angular momentum with time,
as more gas starts to move inward. For $t = 365.9$\,Myr,
the gas inside $\sim 10$\,pc radius from the seed exhibits a nearly maximal allowable rotation, 
$j_{\rm gas}\sim j_{\rm c}$ --- a clear signature that the disk completely dominates the 
kinematics in this region.

We now turn to the main mechanism of angular momentum loss by the gas. If angular
momentum were conserved during the collapse, the gas within the dominant DM halo
potential would be able to collapse only by a factor of $\sim 10$ before forming a disk
\citep[e.g.,][and refs. therein]{Shlosman:13}. However, the shapes of growing DM halos are inherently
triaxial, as seen in virtually all numerical simulations
\citep[e.g.,][]{Allgood.etal:06,Berentzen.etal:06}. The gravitational torques which result
from such halos remove angular momentum from the gas and allow the
collapse to proceed \citep[e.g.,][]{Berentzen.Shlosman:06}. This happens even in regions 
where the gas has decoupled from
the DM, because for non-axisymmetric mass distributions, the external mass can
impose torques on smaller radii.

\begin{figure*}
\centerline{
  \includegraphics[width=0.36\textwidth,angle=0] {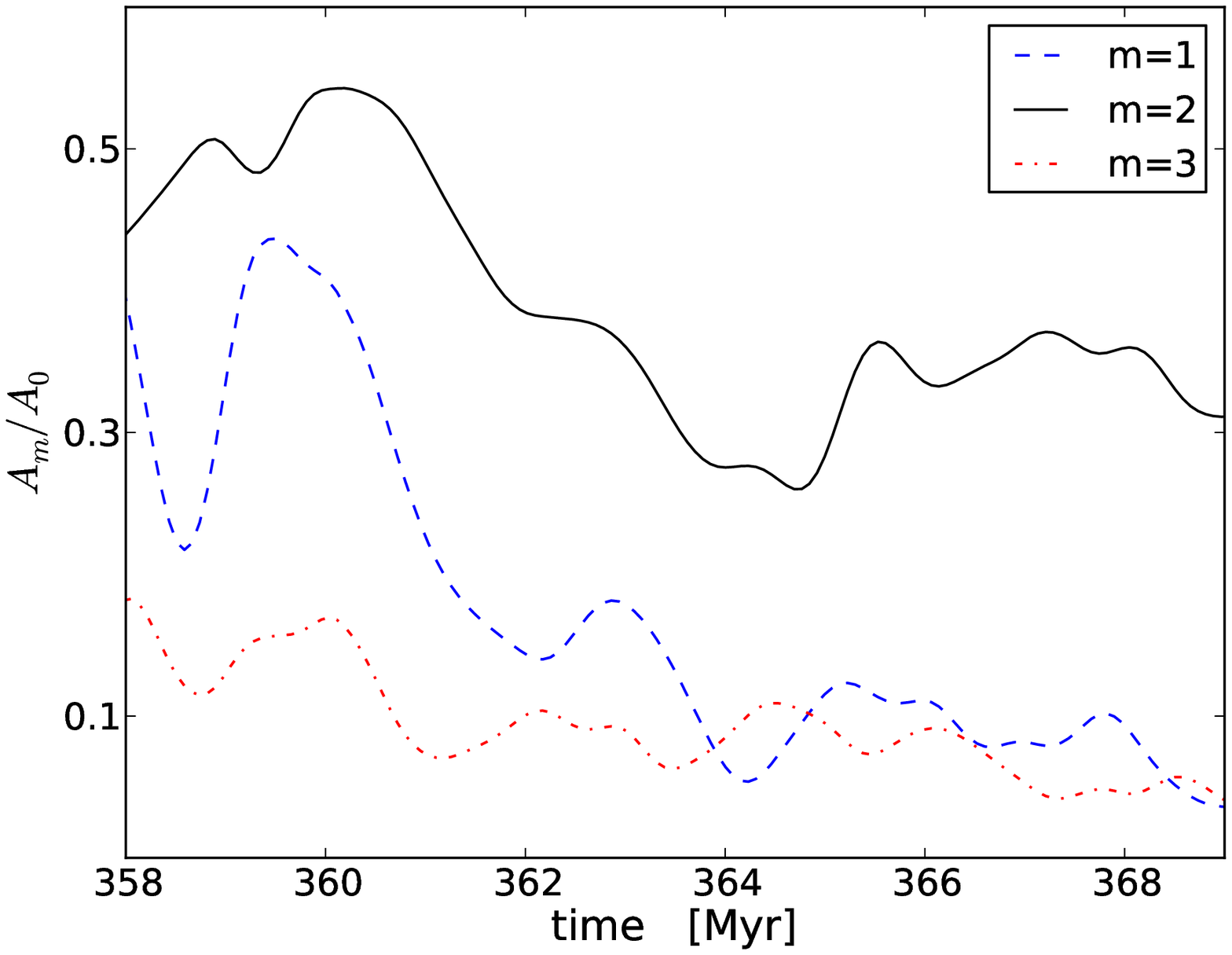}
  \includegraphics[width=0.34\textwidth,angle=0] {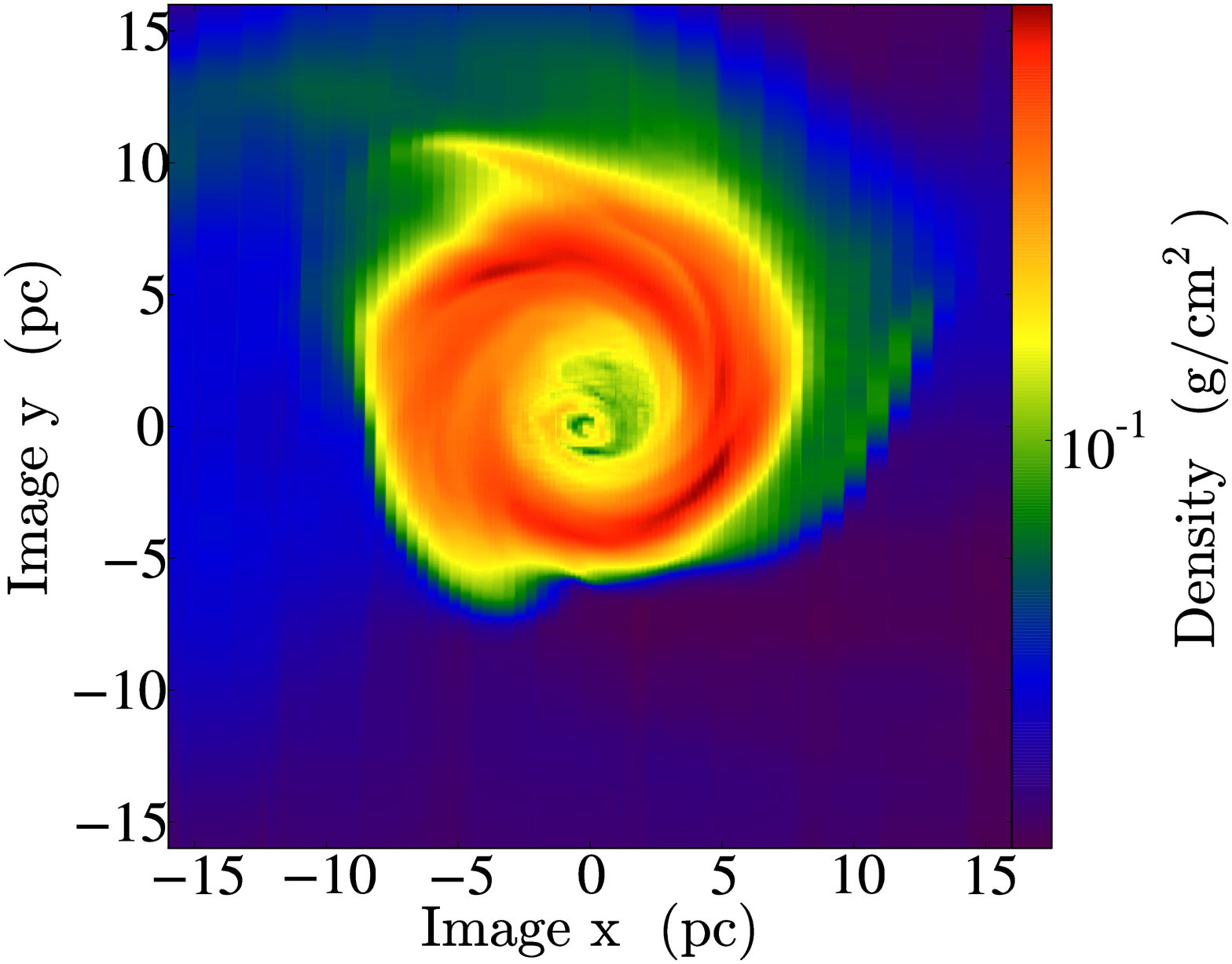}
  \includegraphics[width=0.36\textwidth,angle=0] {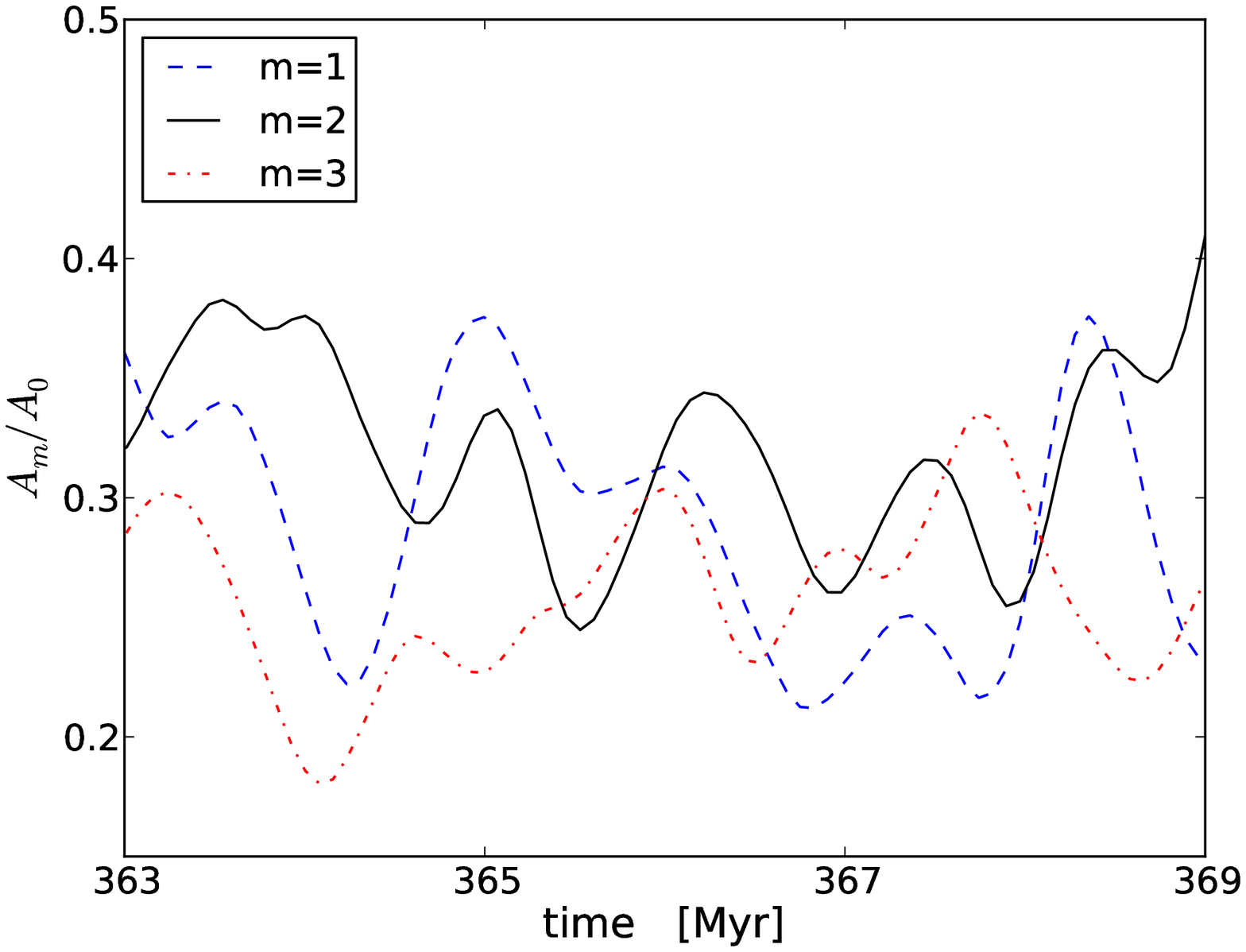}
}
\caption{
Evolution of the nonaxisymmetric modes on progressively smaller spatial scales --- in 
the region dominated by the filaments, and in the outer and inner disks.
{\it Left:} Evolution of the Fourier gas density modes $m=1$, 2, and 3, normalized to the $m=0$ mode 
amplitude for run L12 in the penetrating filaments region. The amplitudes are measured within the
cylindrical annulus defined by $r\sim 10-50$\,pc and thickness $\Delta z = 10$\,pc. 
{\it Middle:} Face-on outer disk and inclined inner disk at the end of the 
simulation, $t = 369.12$\,Myr. Note the effect of the outer filaments penetrating the disk and 
driving $m=2$ and higher modes in the form of spirals.
{\it Right:} Evolution of the Fourier gas density modes $m=1$, 2, and 3, normalized to the $m=0$
mode amplitude for run L12 in the face-on inner disk. The amplitudes are measured
within the cylindrical annulus defined by $r\sim 0.5-2$\,pc and thickness $\Delta z = 0.5$\,pc. 
}
\label{fig:3Mode}
\end{figure*}

We have verified the importance of gravitational torques on large and small spatial scales
in cosmological simulation of direct collapse \citep[][]{Choi.etal:15}. On larger scales,
$r\sim 10-50$\,pc, we have extended this analysis for another $\sim 10$\,Myr, in the
presence of the massive central seed (Fig.\,\ref{fig:3Mode}, left frame). A careful inspection of
this region shows that the high amplitude of the
$m=2$ mode has its origin in a pair of dominant filaments, which feed the outer disk at a
high mass accretion rate. So, indeed, the development of the $m=2$ mode with some non-negligible
amplitude is expected here. Prior to 
the formation of the central seed, the gas on 
these scales responds to the torques from the DM halo, and its response is that of a
self-gravitating fluid. However, after $t\sim
362$\,Myr, the radius of influence of the seed is about $r_{\rm infl}\sim 43\,M_{\rm seed,6} v_{\rm
vir,10}^{-2}$\,pc, where $M_{\rm seed,6}$ is the seed mass in units of $10^6\,M_\odot$, and
$v_{\rm vir,10}$ is the DM halo virial velocity in units of $10\,{\rm km\,s^{-1}}$. Hence,
almost immediately after its formation, the central seed dominates the dynamics of the region
hosting the disks. After this time, the gas response to the external torques, both
hydrodynamical and gravitational, is that of a non-self-gravitating fluid.

The morphological evolution of the outer disk is, therefore, dominated by the mass influx
along the filaments. Without this perturbation, the disk would become largely axisymmetric,
but as the deprojected image of the face-on outer disk shows (Fig.\,\ref{fig:3Mode}, middle frame), 
it is dominated by spiral modes, from $m=2$ to higher harmonics, with substantial amplitudes.
The inner disk, within a cylindrical annulus  
$0.5\ltorder r\ltorder 2$\,pc, and thickness $\Delta z = 0.5$\,pc, exhibits a lower
amplitude $m=2$ and 1 responses (Fig.\,\ref{fig:3Mode}, right frame), and is affected by the 
filaments to a smaller  extent.
Note that the misalignment of the two disks itself can in principle produce gravitational torques 
and induce $m=2$ density modes, albeit heavily diluted by the central seed. 

Thus, the self-gravity of both disks
is severely diluted by the gravity of the massive seed in the center. The outer disk is  perturbed
by the pair of external filaments which maintains the $m=2$ symmetry there. The inflow
rate that reaches the inner disk is much smaller, comes mostly from the outer disk (with infall along
the rotation axis), and 
appears more axially symmetrized.

How stable is the configuration of misaligned disks and why did it form in the first place? 
The answer to the latter question is that the misaligned disks form in response to the
variable angular momentum axis in the accretion flow. Future simulations will show whether
this configuration is limited to a pair of disks, or whether additional misaligned rotating
flows can form at larger radii. The reason for its apparent
stability is found in the two filaments which feed the outer disk. By the end of the simulation,
$t\sim 369.12$\,Myr, these filaments have been substantially degraded, and so one can expect that
the disks will re-align themselves. But gravitational torques between the disks are severely
diluted by the central seed, which helps to prolong the life of this configuration. Of course, 
continuing collapse will add additional mass
to the region, and one cannot predict a priori what will be the direction of its angular momentum
vector.

To summarize, the massive seed fundamentally transforms the character of the accretion flow,
to a large extent independently of the numerical resolution, once the radius of influence of
this seed is resolved. Most importantly, as long as it 
dominates gravitationally, the massive seed dilutes the self-gravity of the gas, damps 
fragmentation, decreases
substantially the angular momentum transfer away from the gas, and, therefore, leads to the 
formation of a disk(s) in its vicinity. The scale chosen to break the self-similar collapse in our runs, 
given by the condition to create sink particles, determines the
position of the disky flow at $\sim 0.01-10$\,pc. Increasing or decreasing this scale would
move the disk inward or outward, respectively, but the appearance of such a flow
is generic.

\subsection{Evolution of turbulent motions in the accretion flow}
\label{sec:turbu}

It is more difficult to recognize the two-disk system in the vorticity magnitude slices displayed
in Figure\,\ref{fig:Vorticity}. Nevertheless, the face-on and edge-on inner disk appears prominent 
in the $yz$
and $xz$ projection planes in the bottom frames. For the edge-on projections of this disk one
observes accretion flows that are directed along its rotation axis. These flows appear to be highly
turbulent. Generally, the vorticity increases toward the center, as the vorticity
maps show on various spatial scales. The flow is turbulent both
away from the outer disk and in its vicinity, though closer to the outer disk mid-plane the 
turbulence decays. Further analysis of the developing turbulence in the collapsing flow
will be presented elsewhere.

\section{Discussion}
\label{sec:discuss}

We have used cosmological zoom-in simulations with the AMR Enzo code to study the long-term evolution of 
direct collapse that can lead to the
formation of SMBH seeds at high redshifts. To circumvent the Courant condition on the
timestep we have invoked the sink particle method.
Sink particles are introduced when the refinement criterion has been violated
and the resolution becomes insufficient at the highest refinement level (see Section\,\ref{sec:method}). 
The innermost accretion flow on scales of $\ltorder 1$\,pc
forms a number of sink particles after $\sim 361.5$\,Myr. They have
been permitted to grow by accretion and merging. By comparing various
spatial resolutions, we find that allowing for a maximum of 12 refinements,
corresponding to 0.1\,pc resolution, is sufficient to obtain the convergence in the growth rate 
of the central sink particle, i.e., this growth rate remains unchanged when the resolution
is increased. Under these conditions, we have been able to prolong
the evolution of the collapsing gas by another 10\,Myr, compared to \citet{Choi.etal:15},
who stopped the simulations at $t=360.13$\,Myr ($z\sim 12$) after the Big Bang. 
Our main results show that, 

\begin{itemize}

\item The masses of these sink 
particles have remained below $\sim 10^3\,M_\odot$, with the exception of the central
particle which grows rapidly to $M_{\rm seed}\sim 2\times 10^6\,M_\odot$ in less than $\sim 2$\,Myr,
due to gas accretion and merging with other particles. 

\item This growth coincides with
the cessation of gravitational collapse on sub-pc scales and the formation of two misaligned 
gas disks --- the inner one after $\sim 361.5$\,Myr, and the outer one after $\sim 363.7$\,Myr. 

\item Formation of misaligned disks is related to the variability of the angular momentum direction in 
the region.

\item As we discuss below (see also Section\,\ref{sec:disk}),
the appearance of the massive object (hereafter central seed) in the center is responsible for the 
cutoff of collapse. 

\item On larger scales, however, the collapse continues unimpeded. At this point,
we make no assumptions about the nature of this massive object, but note that its rapid
formation confirms the feasibility of this path to an SMBH.

\end{itemize}

We have analyzed the co-evolution of the flow with the growth of the central seed, 
on scales comparable to the size of the forming disk. For this purpose, we have
computed the evolution of Fourier modes on scales $\sim 0.5-2$\,pc in the inner
disk, $2.5-10$\,pc in the outer disk, and  $10-50$\,pc outside the disk system. The low $m=1$
to 3 mode amplitudes remain moderate, in the range of $0.2-0.4$ (relative to $m=0$) in the inner 
disk, and the outer disk 
is dominated by $m=2$ spiral modes. They do not show any additional growth which can be
expected if the fluid is self-gravitating.
In both disks, we estimate that gas self-gravity is not important in comparison
with the gravitational acceleration due to the massive central seed, and show that the fragmentation
in the disks is damped due to the sharp increase of the Jeans masses there, after the formation
of the massive seed (Section\,\ref{sec:disk}). In other words, the disks lie within the Roche limit of the
central seed.
However, the inflow into the outer disk is dominated by a pair of filaments
which maintain the $m=2$ perturbation for a long time, while shear tends to axisymmetrize it. The mass 
inflow there is high,
${\dot M}\sim 0.3\,M_\odot\,{\rm yr^{-1}}$. The inner disk is fed at a much
smaller rate, ${\dot M}\ltorder 10^{-2}\,M_\odot\,{\rm yr^{-1}}$, by the outer disk.
Under these conditions of severely diluted self-gravity, mode-coupling does not operate
and the non-axisymmetric (and mostly $m=2$) modes 
quickly saturate \citep[e.g.,][]{Christodoulou.etal:95}.

Because the innermost accretion rate is substantially below that on larger spatial scales, the gas
will accumulate in the disks, as discussed in section\,\ref{sec:disk} and shown in 
Figure\,\ref{fig:masses}. We expect
that this mass growth can lead to some local gravitational instabilities in the disks, but not
to the resumption of a massive accretion onto the central seed. 

A number of important issues are related to this evolution.
Probably the most interesting one is how the collapse proceeds on smaller scales --- this has
been deliberately ignored
in the current work, with the introduction of the seed particle algorithm. We anticipate that the 
cessation of the accretion
process observed here justifies our conclusion that the characteristic mass of the forming
SMBH seed is $M_\bullet\sim 10^6\,M_\odot$. 

Another issue is related to the fate of the inclined disks
configuration, which formed as a result of gas flowing into the region with a variable orientation of 
angular momentum. Most probably this configuration will be destroyed by the same process that created
it in the first place. In any case, if the SMBH seed forms before the next spike in the accretion rate,
the radiation and/or mechanical feedback could already modify substantially the conditions in the 
inflowing gas.

The presence of the massive central object, can be favorable for the growth of the $m=1$ mode 
in the surrounding accretion disk, when the disk mass is {\it not} negligible compared to the massive
object --- hence when disk self-gravity is not severely diluted by the object
\citep[e.g.,][]{Adams.etal:89}. Furthermore, for {\it stellar} disks, which are dominated by the
central object and hence reside in the Keplerian potential, the $m=1$ mode, when triggered, can
be long-lived, due to the absence of precession in this potential
\citep[][]{Tremaine:95}. Under these conditions, the individual stellar orbits can be weakly
``glued'' by their gravitational attraction and can maintain this non-axisymmetric mode.
This mode has been suggested to be responsible for the mass accretion rate onto the SMBH under
certain conditions \citep[][]{Hopkins.Quataert:10}.

However, one expects that the $m=1$ mode will not be able to grow when strong time-dependent
perturbations are present, which destroy the orbital correlations in the Keplerian gaseous disk,
i.e., in a disk with the ratio of the central mass-to-disk mass much greater than unity.
We do not observe any substantial growth of $m=1$ mode and any mass accretion flux associated
with this mode.

Fast growth of the central object will have dynamical consequences for
the collapsing flow. Among these, the most important appears to be the {\it inability} of the flow to get
rid of its angular momentum. Gravitational torques from the DM serve as the main mechanism
extracting angular momentum from the gas, but close to the massive central seed, the
torques are diluted by the gravitational monopole and become inefficient. The gas
acquires rotational support and settles in a disk-like configuration around
the center. Its fragmentation is damped by the tidal forces of the central seed in the presence of
the floor temperature of atomic gas.

Damping of the nonaxisymmetric instabilities in the disk leads to the inability
of the gas to transfer its angular momentum to the outer
gas or to the surrounding DM. This results in the accumulation of gas in this region and
the growth of the disk. In fact, the rapid growth of the central object is the prime cause
of the formation of the disks on these scales and of their lack of fragmentation.
After the initial period of rapid growth, the massive central object continues to grow
but at a much slower pace, by a factor of $\sim 2-3$ (Fig.\,\ref{fig:SinkEvolution}). 

The corollary of this evolution is the formation of a massive object in the
center of the direct collapse region. The resolution scale in our model is chosen
somewhat arbitrarily, just to allow for the second stage of the collapse to be resolved
spatially. But our results bring up an important issue: how does the nature of the accretion
flow changes with the formation and growth of the central massive object? The resulting
formation of the gaseous disk around this object is a direct consequence of dramatically
reduced efficiency of angular momentum transfer within the radius of influence of the
massive center, $R_{\rm infl}$ (Section\,\ref{sec:ang_mom}). Of course the compactness parameter of such 
an object, $M_{\rm seed}/R_{\rm
seed}$, where $R_{\rm seed}$ is the seed size, will determine the inner edge of the forming disk, 
but not the position 
of its outer edge, as long as it is resolved numerically. Here $R_{\rm seed}$  
is the size of the seed or the associated numerical resolution.

We have also assumed that the molecular gas, H$_2$, is not present during direct collapse, and, therefore,
the floor temperature of the atomic gas lies around a several thousand degrees. We justify this by
the presence of the Lyman-Werner continuum which originates in nearby stellar populations 
\citep[e.g.,][]{Agarwal.Khochfar:15} or in situ \citep[][]{Choi.etal:13}.

Another important issue is our neglect of the radiation feedback in the region of the disk.
To address this, it is crucial to know the nature of the massive object, and the source of the radiation
field. If this object is an SMS fueled by the thermonuclear reactions in the core and accretion
energy in the envelope \citep[e.g.,][]{Begelman:10}, this will specify 
the effective temperature, spectral energy distribution and geometry of the radiation field.
Alternatively, if the SMS stage is bypassed (see Section\,1), the rate of locally produced energy will
be very different and the radiation field could become anisotropic, with most of it escaping in the
preferred direction, i.e., the rotation axis. We also note that very large accretion rates
encountered during direct collapse can make the radiation feedback much less efficient
dynamically.

\section*{Acknowledgments}
We thank Takashi Hosokawa for interesting and insightful discussions, and thank the Enzo and YT 
support team.  All analysis has been conducted using YT
\citep[][http://yt-project.org/]{Turk.etal:11}. I.S. acknowledges  support from
NSF grant AST-0807760 and from HST/STScI grant AR-12639.01-A. I.S. and K.N. are grateful for support
from International Joint Research Promotion Program at Osaka University.
J.H.C acknowledges support from NASA ATP NNX11AE09G, NSF AST-1009799, and Caltech/JPL SURP Project 
No. 1515294 through the UT Austin (P.I. Paul Shapiro). M.C.B. acknowledges
support from the NSF under AST-1411879. K.N. acknowledges the partial support by JSPS KAKENHI 
Grant Number 26247022. Support for
HST/STScI AR-12639.01-A was provided by NASA through a grant from the STScI, which is 
operated by the AURA, Inc., under NASA contract NAS5-26555.

\bibliographystyle{mn}
\bibliography{MyRef}


\end{document}